\documentclass[aps,prc,twocolumn,superscriptaddress,floatfix,nofootinbib,longbibliography]{revtex4-2}

\usepackage[utf8]{inputenc}
\usepackage[english]{babel}
\usepackage[dvipsnames]{xcolor}
\usepackage{amssymb,amsthm,amsmath,amstext,amsbsy,amsopn}
\usepackage{bbm}
\usepackage{graphicx}
\usepackage{isotope}
\usepackage{tabularx}
\usepackage{siunitx}

\newcolumntype{Y}{>{\centering\arraybackslash}X}
\usepackage{float}
\usepackage{orcidlink}
\usepackage{hyperref}
\hypersetup{
    colorlinks = true,
    citecolor  = blue,
    linkcolor  = blue,
    urlcolor  = blue
}

\newcommand{\mbptorder}{k}
\newcommand{\Eref}{E_\textnormal{HF}}
\newcommand{\Erefstat}{E_\textnormal{ref}}
\newcommand{\prob}[1]{\textnormal{pr}\left(#1\right)}
\newcommand{\normal}[2]{\mathcal{N}\left(#1, #2\right)}
\newcommand{\ig}{\mathcal{IG}}
\newcommand{\gbar}{\bar{\gamma}}
\newcommand{\alphag}{\alpha}
\newcommand{\betag}{\beta}
\newcommand{\NN}{NN}
\newcommand{\NNN}{3N}
\newcommand{\resolvent}{R_\text{RS}}

\newcommand{\la}{\langle}
\newcommand{\ra}{\rangle}

\newcommand{\MeV}{\ensuremath{\text{MeV}}}

\newcommand{\elem}[2]{\ensuremath{^{#1}\text{#2}}}

\newcommand{\magicint}{1.8/2.0 (EM)}
\newcommand{\deltago}{$\Delta$N$^2$LO$_\text{GO}$}
\newcommand{\emarthuis}{1.8/2.0 (EM7.5)}

\newcommand{\nnlosat}{N$^2$LO$_\text{sat}$}

\newcommand{\ai}{\textit{ab initio}}
\newcommand{\eg}{e.g.}
\newcommand{\ie}{i.e.}
\newcommand{\pte}[1]{\ensuremath{E^{(#1)}}}

\begin{document}

\allowdisplaybreaks

\author{I.~Svensson \orcidlink{0000-0002-9211-5555}}
\email{isak.svensson@tu-darmstadt.de}
\affiliation{Technische Universit\"at Darmstadt, Department of Physics, 64289 Darmstadt, Germany}
\affiliation{ExtreMe Matter Institute EMMI, GSI Helmholtzzentrum f\"ur Schwerionenforschung GmbH, 64291 Darmstadt, Germany}
\affiliation{Max-Planck-Institut f\"ur Kernphysik, Saupfercheckweg 1, 69117 Heidelberg, Germany}

\author{A.~Tichai \orcidlink{0000-0002-0618-0685}}
\email{alexander.tichai@tu-darmstadt.de} 
\affiliation{Technische Universit\"at Darmstadt, Department of Physics, 64289 Darmstadt, Germany}
\affiliation{ExtreMe Matter Institute EMMI, GSI Helmholtzzentrum f\"ur Schwerionenforschung GmbH, 64291 Darmstadt, Germany}
\affiliation{Max-Planck-Institut f\"ur Kernphysik, Saupfercheckweg 1, 69117 Heidelberg, Germany}

\author{K.~Hebeler \orcidlink{0000-0003-0640-1801}}
\email{kai.hebeler@physik.tu-darmstadt.de}
\affiliation{Technische Universit\"at Darmstadt, Department of Physics, 64289 Darmstadt, Germany}
\affiliation{ExtreMe Matter Institute EMMI, GSI Helmholtzzentrum f\"ur Schwerionenforschung GmbH, 64291 Darmstadt, Germany}
\affiliation{Max-Planck-Institut f\"ur Kernphysik, Saupfercheckweg 1, 69117 Heidelberg, Germany}

\author{A.~Schwenk \orcidlink{0000-0001-8027-4076}}
\email{schwenk@physik.tu-darmstadt.de}
\affiliation{Technische Universit\"at Darmstadt, Department of Physics, 64289 Darmstadt, Germany}
\affiliation{ExtreMe Matter Institute EMMI, GSI Helmholtzzentrum f\"ur Schwerionenforschung GmbH, 64291 Darmstadt, Germany}
\affiliation{Max-Planck-Institut f\"ur Kernphysik, Saupfercheckweg 1, 69117 Heidelberg, Germany}

\title{Bayesian approach for many-body uncertainties in nuclear structure: \\ Many-body perturbation theory for finite nuclei}

\begin{abstract}
A comprehensive assessment of theoretical uncertainties defines an important frontier in nuclear structure research. Ideally, theory predictions include uncertainty estimates that take into account truncation effects from both the interactions and the many-body expansion. While the uncertainties from the expansion of the interactions within effective field theories have been studied systematically using Bayesian methods, many-body truncations are usually addressed by expert assessment. In this work we use a Bayesian framework to study many-body uncertainties within many-body perturbation theory applied to finite nuclei. Our framework is applied to a broad range of nuclei across the nuclear chart calculated from two- and three-nucleon interactions based on chiral effective field theory. These developments represent a step towards a more complete and systematic quantification of uncertainties in \ai{} calculations of nuclei.
\end{abstract}
 
\maketitle

\section{Introduction}

The \ai{} description of nuclear many-body systems is advancing across the nuclear chart and nuclear matter~\cite{Herg20review,Hebe203NF}.
Applications have been pushing predictions to the limits of stability~\cite{Stroberg2021,Tichai2024bcc}, to heavy nuclei~\cite{Hu2021lead,Miyagi:2023zvv,Arthuis2024}, complex deformed nuclei~\cite{Hagen2022PCC,Frosini2021mrII,Sun:2024iht}, and nuclei relevant for beyond-standard-model physics~\cite{Cirigliano:2022rmf,Belley:2023lec,Door2025ytterbium}.
For nuclear matter, advances include calculations up to high order and using different many-body approaches~\cite{Dris17MCshort,Marino:2024tfp,Tews:2024owl,Armstrong:2025tza} as well as calculations for arbitrary proton fraction and temperatures~\cite{Keller2023}.

The quantification of theoretical uncertainties has become an integral part of nuclear physics research in recent years. First-principles calculations have the advantage of allowing the assessment of truncation effects in a systematic way.
Modern \ai{} calculations usually start from nuclear forces based on chiral effective field theory (EFT)~\cite{Epel09RMP,Mach11PR}. The resulting interactions are then combined with systematically improvable many-body approximations for the solution of the Schr\"odinger equation,
\begin{align}
    H | \Psi_n \ra = E_n |\Psi_n\ra \, ,
\end{align}
where $H$ denotes the nuclear Hamiltonian, and $|\Psi_n\ra$ the eigenstate with associated energy $E_n$.
The nuclear Hamiltonian is given by
\begin{align}
    H = T + V_\text{\NN} + V_\text{\NNN} + ... \, ,
\end{align}
where $T$ denotes the (intrinsic) kinetic energy, and $V_\text{\NN}$ and $V_\text{\NNN}$ are two- (\NN{}) and three-nucleon (\NNN{}) interactions, respectively.

Over the past decades different many-body frameworks have been developed that allow to solve the Schr\"odinger equation in approximate yet systematically improvable ways at tractable computational cost. Methods include  coupled-cluster (CC) theory~\cite{Hage14RPP}, the in-medium similarity renormalization group (IMSRG)~\cite{Herg16PR,Stroberg2019}, Green's function approaches~\cite{Barb17SCGFlectnote,Soma20SCGF}, and many-body perturbation theory (MBPT)~\cite{Tichai2020review}.
For lighter systems ($A \lesssim 20$), quantum Monte Carlo methods~\cite{Carl15RMP} and large-scale diagonalizations like the no-core shell model (NCSM)~\cite{Barr13PPNP,Roth09ImTr} are available.

The main sources of theoretical uncertainties in \ai{} calculations are due to
\begin{enumerate}
    \item[$i)$] neglected higher-order contributions in the EFT power counting,
    \item[$ii)$] different fitting strategies and associated experimental uncertainties in the inference of the low-energy couplings in $H$,
    \item[$iii)$] truncations in the many-body expansion when solving the Schr\"odinger equation, and
    \item[$iv)$] model-space truncations from the finite size of the underlying computational basis.
\end{enumerate} 
Clearly, a comprehensive uncertainty quantification (UQ) must encompass (at least) these contributions. Recent years have seen considerable efforts in assessing interaction uncertainties by performing order-by-order calculations in the power-counting and estimating their uncertainties from EFT arguments~\cite{Furn15uncert,Mele17bayerror,Wesolowski:2018lzj,Hopp19medmass,Melendez:2019izc,Huth19chiralfam,Drischler:2020hwi,Hu2021lead,Maris:2020qne,Wesolowski:2021cni,Svensson:2021lzs,Svensson:2022kkj,Jiang:2022tzf,Keller2023,Svensson:2023twt,Millican:2024yuz}. However, these efforts remain unparalleled for the many-body expansion itself. All attempts so far relied on estimates of the effects of higher-order contributions based on expert assessment driven by experience. The goal of this work is the design of error models for many-body expansions. 
In this work we focus on the simplest case of such an expansion, \ie{}, MBPT.
The developed scheme will be benchmarked for a broad range of nuclei and explored in nuclear matter. We will also study the impact of using different chiral interactions.
While model-space truncations can induce sizable uncertainties in heavier systems, they do not pose the dominant source of uncertainty for the present study and we do not take them into account here.

The structure of this work is as follows: In Sec.~\ref{sec:mbpt} we briefly review the basics of MBPT and introduce the error model. Section~\ref{sec:stat} discusses the Bayesian framework for the determination of the theory uncertainties.
In Sec.~\ref{sec:nuclei} we apply our UQ framework to MBPT calculations of a broad range of closed-shell nuclei with some exploratory applications to nuclear matter.
Finally, we conclude in Sec.~\ref{sec:outlook}. The code and data needed to reproduce our results are available as a Zenodo~\cite{zenodorepo} and Github\footnote{\url{https://github.com/svisak/manybody_uncertainties}} repository.

\section{Many-body perturbation theory}
\label{sec:mbpt}

\subsection{Partitioning and reference state}

The main goal of this work is quantification of many-body uncertainties in approximate solutions of the Schr\"odinger equation obtained within MBPT~\cite{Shav09MBmethod}.

In MBPT, the Hamiltonian is partitioned according to
\begin{align}
H_\lambda = H_0 + \lambda H_1 \, ,    
\end{align}
where $H_0$ denotes the unperturbed part and $H_1 \equiv H - H_0$ the perturbation. 
The quantity $\lambda$ is an auxiliary parameter used in the expansion that is finally set to unity.
$H_0$ must in practice be amenable to a simple (numerical) solution
\begin{align}
    H_0 | \Phi_n \ra  = E^{(0)}_n | \Phi_n \ra \, .
\end{align}
The unperturbed ground state $|\Phi \ra \equiv |\Phi_0\ra$ serves as the zeroth-order reference state for basis expansion methods.
In closed-shell nuclei the reference state is commonly obtained from a variational mean-field calculation, \eg{}, using spherical Hartree-Fock (HF) theory, giving rise to a single Slater determinant
\begin{align}
    | \Phi \ra = \prod_{i}^A c^\dagger_i | 0 \ra \, ,
\end{align}
where $c^\dagger_i$ denotes the single-particle creation operator acting on the physical vacuum $|0\ra$.

Following HF theory every single-particle state $|\varphi_p\ra = c^\dagger_p |0\ra$ comes with an associated single-particle energy $\epsilon_p$ that corresponds to the eigenvalues of the self-consistently optimized mean-field Hamiltonian
\begin{align}
    h | \varphi_p \ra = \epsilon_p | \varphi_p \ra \, .
\end{align}
The shell gap is then defined as the difference between the energetically lowest virtual orbital and the energetically highest occupied orbital
\begin{align}
    \Delta_\text{shell} = 
    \min_{p \, \text{virtual}} \epsilon_p - \max_{p \, \text{occupied}} \epsilon_p \, .
\end{align}
This quantity defines the minimum energy needed to excite the system and should be positive in practice.

\subsection{Power-series expansion}

The MBPT ground-state energy is parametrized through a Taylor expansion,
\begin{align}
    E_0 &= \sum_{k=0}^\infty \pte{k} \, \lambda^k \, , 
\end{align}
where $\pte{k}$ defines the $k$th-order energy corrections and $E_0$ the exact ground-state energy. 
The final ground-state energy is evaluated at the physical point $\lambda=1$. The leading-order corrections correspond to the reference state expectation value 
\begin{align}
    \Eref = \la \Phi | H | \Phi \ra  = \pte{0} + \pte{1} = E^{(0)}_0 \,,
\end{align}
such that the HF energy includes contributions up to first order in the interactions, and MBPT corrections on top of HF start at second order. 
Energy corrections are practically evaluated in terms of the Goldstone expansion,
\begin{align}
    \Delta E = \la \Phi | \sum_{k=1}^\infty H [ \resolvent H_1 ]^{k} | \Phi \ra_c \, ,
    \label{eq:goldstone}
\end{align}
where $\Delta E = E - \Eref$ denotes the correlation energy and the lower index $c$ indicates the connected character of the expansion~\cite{Gold57mbpt}.
The operator $\resolvent$ denotes the so-called (Rayleigh-Schr\"odinger) resolvent
\begin{align}
    \resolvent \equiv \sum_{k \neq 0} \frac{|\Phi_k\ra\la\Phi_k|}{E^{(0)}_0 - E^{(0)}_k} \, ,
\end{align}
where the sum runs over all states but the reference state.
The use of a Rayleigh-Schr\"odinger resolvent results in a manifestly size-extensive expansion, \ie{} an expansion whose relative error is independent of the system size.

Expanding the Goldstone formula gives rise to the second-order energy correction written in configuration space,
\begin{align}
    \pte{2} = \sum_k \frac{\la \Phi | H_1 | \Phi_k \ra \la \Phi_k | H_1 | \Phi\ra}{E^{(0)}_0 - E^{(0)}_k} \, .
    \label{eq:mbpt2}
\end{align}
MBPT expressions are evaluated in many-body systems through the use of Wick's theorem that allows for a rewriting in terms of second-quantized many-body operators, leading to a second-order correction
\begin{align}
    E^{(2)} = \frac{1}{4} \sum_{abij} \frac{H_{abij} H_{ijab}}{\epsilon^{ab}_{ij}} \, ,
\end{align}
where $H_{abij}$ denote two-body matrix elements of the normal-ordered Hamiltonian and the short-hand notation 
\begin{align}
\epsilon^{ab}_{ij} \equiv \epsilon_i + \epsilon_j - \epsilon_a - \epsilon_b
\end{align}
involves the HF single-particle energies $\epsilon_p$.
Here we follow the index convention where occupied states (holes) are denoted by $i,j,...$, whereas unoccupied states (particles) are denoted by $a,b,...$.

The derivation of algebraic expressions is aided by the use of diagrammatic techniques~\cite{Shav09MBmethod}.
The evaluation of higher-order corrections becomes increasingly more challenging as the number of diagrams grows exponentially and the computational cost of their individual evaluation increases substantially. Hence MBPT applications in nuclear physics are limited to order $k=4$ in selected applications~\cite{Miyagi2021}, while routinely only corrections up to third order are evaluated, as is done in this study.

\subsection{Convergence of MBPT expansion}

As can be seen from the Goldstone expansion, Eq.~\eqref{eq:goldstone}, higher-order energy corrections are driven by the operator $(\resolvent H_1)$.
Consequently, the rate of convergence critically depends on properties of the resolvent and perturbation operator, where the size of the energy denominator of the resolvent operator is directly linked to the shell gap, and the size of the coupling to virtual states is linked to the action of the perturbation operator $H_1$.
The shell gap itself is strongly system dependent and may be significantly weakened if a nucleus' closed-shell character is less pronounced, \eg{}, in the case of structurally more complex nuclei such as \elem{12}{C}. Doubly magic nuclei with major shell closures at $N, Z=8,20,28,50,82$ typically feature pronounced shell gaps and hence allow for an efficient treatment from HF-based expansions.
The coupling to virtual states on the other hand is interaction-specific and can---for the same nucleus---result in a very different convergence patterns of MBPT. In particular ``harder'' interactions, \eg{}, at a higher resolution scale, typically yield an enhancement of higher-order MBPT corrections and hence a deterioration of the speed of convergence up to a point where the perturbative expansion may quickly become divergent~\cite{Tich16HFMBPT}.
Thus, the behavior of an MBPT expansion is the result of a complex interplay between the choice of the reference state and properties of the interaction.

\begin{figure}[t!]
    \centering
    \includegraphics[width=\columnwidth]{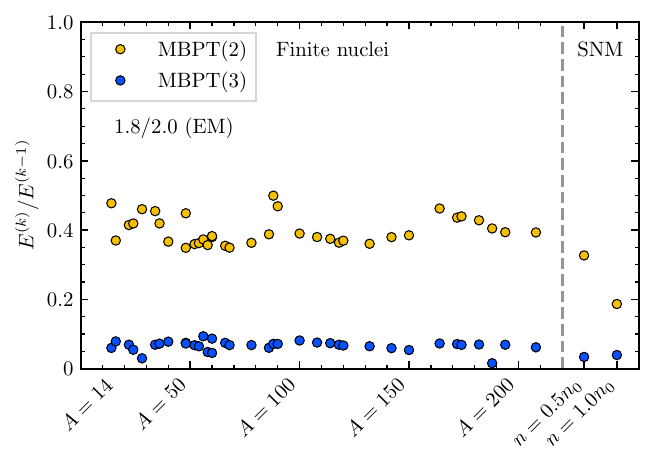}
    \caption{Ratios $\pte{2}/\Eref$ and $\pte{3}/\pte{2}$ for our set of nuclei from Ref.~\cite{Arthuis2024} as well as symmetric nuclear matter at two representative densities from Ref.~\cite{Alp:2025wjn}. $n_0 = 0.16$\,fm$^{-3}$ denotes the nuclear saturation density. All shown results have been computed using the \magicint{} interaction~\cite{Hebe11fits}.}
    \label{fig:mbptratios}
\end{figure}

We therefore motivate our UQ analysis by displaying the energy corrections for the \magicint{} interaction~\cite{Hebe11fits} for a broad range of doubly closed-shell nuclei from Ref.~\cite{Arthuis2024} as well as for symmetric nuclear matter at two representative densities from Ref.~\cite{Alp:2025wjn} (details are introduced in Sec.~\ref{sec:nuclei}).
As shown in Fig.~\ref{fig:mbptratios} the values of $\pte{2}/\Eref \approx 0.4$ are significantly larger than the corresponding values of $\pte{3}/\pte{2} \approx 0.1$.
We note that the ratio $\pte{3}/\pte{2}$ is unnaturally small. In fact, fourth-order results for \elem{16}{O} revealed a ratio of $\pte{4}/\pte{3} = 0.49$, which is comparable to $\pte{2}/\Eref$~\cite{Miyagi25pc}. 
Due to a lack of model-space convergence, it is however not possible to provide reliable results for the ratio $\pte{4}/\pte{3}$ in heavier systems.
Still, all calculated ratios are well below $\pte{k+1}/\pte{k} < 1$ which is expected for a soft interaction like \magicint{} (see Sec.~\ref{sec:limitations} for a counterexample).

Finally, we emphasize the remarkably constant value of the observed MBPT ratios throughout the range of nuclei studied.
This is linked to the size-extensive character of the MBPT corrections at any order, leading to manifestly constant ratios.
The observed small-scale variations with a spread of about $0.05$ are due to finite-size effects linked to, \eg{}, nuclear shell structure.
We note that by this reasoning, one can in principle generate fourth-order pseudo-data based on the value of $\pte{4}/\pte{3}$ in \elem{16}{O}.
However, in this work we do not follow this approach and leave this open for future study.

\section{Statistical framework}
\label{sec:stat}

\subsection{Error model for MBPT expansion}

The proposed model for the MBPT truncation error is inspired by the work of the BUQEYE collaboration on EFT truncation uncertainties~\cite{Furn15uncert,Weso16Bayes,Mele17bayerror,Wesolowski:2018lzj,Melendez:2019izc}. 
We assume that the MBPT expansion follows a power law $\pte{k+1}/\pte{k} \approx R$ with some fixed value $R$ and a set of expansion coefficients $\{\gamma_i\}$ drawn from an underlying (unknown) probability distribution, \ie{}, 
\begin{equation}
    \delta E_0 = \Erefstat \sum_{i=\mbptorder}^\infty \gamma_i R^i \, ,
    \label{eq:truncation_error}
\end{equation}
where $\delta E_0$ is the truncation error of the ground-state energy. Note that this definition means that the second-order MBPT correction is suppressed by (approximately) a factor $R$, the third order by $R^2$, and so on:
\begin{equation}
\begin{split}
    E_0 &= \underbrace{\Erefstat \gamma_0}_{E_\text{HF}} + \underbrace{\Erefstat \gamma_1 R^1}_{\text{MBPT(2)}} + \underbrace{\Erefstat \gamma_2 R^2}_{\text{MBPT(3)}} \\ &\quad + \ldots + \underbrace{\Erefstat \gamma_{k-1} R^{k-1}}_{\text{MBPT(k)}} + \ldots \,.
\end{split}
\end{equation}
The term $\Erefstat$ factors out an overall ground-state energy scale and is set to be $\Erefstat = - 8A \, \MeV$, where $A$ is the mass number. This empirical value is motivated by the typical binding energy per nucleon across the nuclear chart.
The size of $R$ relates to the ``softness'' of the interaction as it (up to small random variations captured by $\gamma_i$) encodes the rate of convergence; in particular, the case of $R > 1$ accounts for the possibility of a divergent expansion and hence a breakdown of the MBPT approach that is implicitly allowed within our approach. 

\subsection{Distribution for expansion parameters $\gamma$}

The expansion coefficients $\gamma_i$ can be re-expressed as
\begin{equation}
    \gamma_i = \frac{\pte{\mbptorder}}{\Erefstat R^i} \quad \text{for} \quad i = k-1 \,,
    \label{eq:compute_gamma}
\end{equation}
which will be used to infer the distributions of $R$ and $\gamma_i$.
Assuming that the $\gamma_i$'s are normally distributed with variance $\gbar^2$,
\begin{equation}
    \prob{\gamma_i} = \normal{0}{\gbar^2} \,,
\end{equation}
leads to a normal distribution for the truncation error
\begin{equation}
    \prob{\delta E_0 | R, \gbar^2} = \normal{0}{\sigma^2} \,,
    \label{eq:truncation_distribution}
\end{equation}
with the variance given by
\begin{equation}
    \sigma^2 = \Erefstat^2 \gbar^2 \sum_{i=\mbptorder}^\infty R^{2i} \,.
    \label{eq:gamma_stddev}
\end{equation}
The assumption of a normal distribution for $\gamma_i$ is in line with the principle of maximum entropy, given that the distribution should have infinite support and we have testable information about its mean and variance; this is analogous to EFT truncation errors~\cite{Furn15uncert}. In this case we have very limited data (effectively 2-3 points) to verify that the assumption of normality holds and properly testing this is a subject for future work.

If the series in Eq.~\eqref{eq:gamma_stddev} is convergent ($R < 1$), we can replace the infinite sum by~\cite{Wesolowski:2018lzj}
\begin{equation}
    \sigma^2 = \Erefstat^2 \gbar^2 \frac{R^{2k}}{1-R^2} \,.
\end{equation}
In the context of EFT truncation errors this expression is sufficient because the series is assumed to be convergent. Here, however, we allow for the possibility of a divergent series, and in cases where $R \geq 1$ we instead set $\sigma^2$ to the largest noninfinite value [\texttt{numpy.finfo(numpy.float64).max}] allowed by NumPy's~\cite{harris2020array} implementation of the normal distribution.

Once $R$ and $\gbar^2$ are known, we can add random samples from Eq.~\eqref{eq:truncation_distribution} to the bare MBPT result to arrive at a posterior predictive distribution (PPD) for the ground-state energy $E_0$.

\begin{figure}[t!]
    \centering
    \includegraphics[width=0.9\columnwidth]{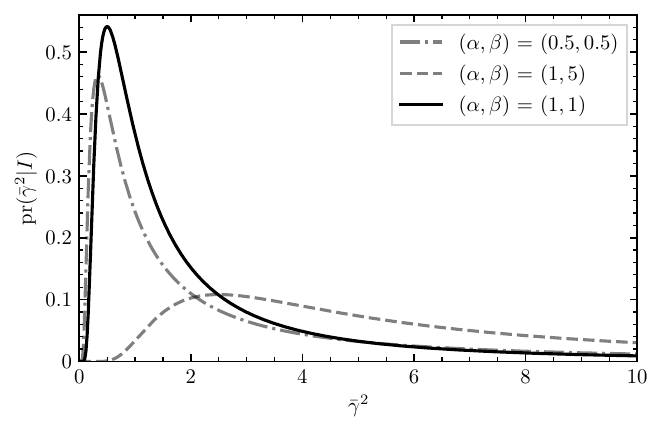}    
    \caption{Conjugate inverse-gamma prior for $\gbar^2$ with hyperparameters given by the legend. Our chosen prior is shown as a solid black line. The two alternative priors, shown as gray dashed and dot-dashed lines, are discussed in App.~\ref{app:priors}.}
    \label{fig:gamma2_prior}
\end{figure}

\subsection{Bayesian inference of truncation-error hyperparameters}

We infer the model parameters $R$ and $\gbar^2$ using a slightly modified version of the procedure introduced in Ref.~\cite{Melendez:2019izc} and applied in Ref.~\cite{Wesolowski:2021cni}, where the analogous parameters $Q$ and $\bar{c}^2$ were inferred in the context of few-nucleon systems. By the product rule of probability theory, the joint probability distribution function for $R$ and $\gbar^2$ can be expressed as
\begin{equation}
    \prob{R, \gbar^2 | \vec{\gamma}, I} = \prob{\gbar^2 | R, \vec{\gamma}, I} \prob{R | \vec{\gamma}, I} \,,
    \label{eq:R_gamma_posterior}
\end{equation}
where the factors on the right-hand side are the Bayesian (marginal) posteriors for $\gbar^2$ and $R$, respectively, given the input data $\vec{\gamma} = \{\gamma_i(R)\}$ computed using Eq.~\eqref{eq:compute_gamma}. It is important to note that while this input data is strongly correlated across the broad range of nuclei, we treat it as uncorrelated in the formalism below. To limit the influence of these correlations we will only use three of the nuclei shown in Fig.~\ref{fig:mbptratios} in the inference of $\prob{R, \gbar^2 | \vec{\gamma}, I}$. A rigorous treatment of correlated data is a subject for future work.

We will now define explicit expressions for the two factors in Eq.~\eqref{eq:R_gamma_posterior}, starting with $\prob{\gbar^2 | R, \vec{\gamma}, I}$. Since the data $\vec{\gamma}$ are assumed to be normally distributed, we place an inverse-gamma ($\ig$) prior with hyperparameters $(\alphag, \betag)$ on its variance $\gbar^2$,
\begin{equation}
    \prob{\gbar^2 | I} = \ig \left(\alphag, \betag\right) = \ig \left(1, 1\right) \,,
\end{equation}
which encodes our prior belief that $\gbar$ is naturally sized. The prior is plotted in Fig.~\ref{fig:gamma2_prior}. The conjugacy of this prior with respect to the normally distributed data results in a posterior that is also an $\ig$ distribution but with updated hyperparameters $(\alphag', \betag')$,
\begin{equation}
    \prob{\gbar^2 | R, \vec{\gamma}, I} = \ig\left(\alphag', \betag'\right) \,.
\end{equation}
The posterior for $\gbar^2$ is conditionally dependent on $R$, whose posterior is discussed below, via Eq.~\eqref{eq:compute_gamma}. The posterior hyperparameters $\alphag'$ and $\betag'$ are obtained analytically using the updating formulas~\cite{Svensson:2022kkj}
\begin{subequations}
\begin{align}
    \alphag' &= \alphag + \frac{N_\text{nuclei} N_\text{orders}}{2} \, , \\
    \betag' &= \betag + \frac{\vec{\gamma}^2}{2} \, ,
\end{align}
\end{subequations}
where $N_\text{nuclei}$ is the number of nuclei and $N_\text{orders}$ the number of orders $\mbptorder$ used in the inference. In other words, the length of $\vec{\gamma}$ is $N_\text{nuclei} \times N_\text{orders}$.

No suitable conjugate prior exists for $R$. For this prior we therefore choose an uninformative uniform prior in the range $R \in (0, 2)$. It also allows for $R > 1$, \ie{}, that the MBPT series could be divergent. The latter is an important difference here compared to EFT truncation errors.
With these priors in place, the posterior for $R$ is given by~\cite{Melendez:2019izc,Wesolowski:2021cni}
\begin{equation}
    \prob{R | \vec{\gamma}, I} \propto \frac{\prob{R | I}}{\left(\frac{\betag'}{\alphag'}\right)^{\alphag'} \prod_{\mbptorder} R^{N_\text{nuclei}(\mbptorder-1)}} \,.
    \label{eq:R_posterior}
\end{equation}
Unlike in Ref.~\cite{Wesolowski:2021cni}, we do not need to account for the additional complication of normalizing Eq.~\eqref{eq:R_posterior} because here the normalization factor does not depend on the quantities we are ultimately interested in. We proceed to sample the two-dimensional joint posterior, Eq.~\eqref{eq:R_gamma_posterior}, using the Markov-chain Monte Carlo (MCMC) sampler \texttt{Emcee}~\cite{Foreman_Mackey_2013}. The computational cost of the MCMC sampling is insignificant and well-converged chains are obtained in a matter of minutes on current hardware.

\section{Application to finite nuclei}
\label{sec:nuclei}

\subsection{Computational setup}

We start by investigating many-body uncertainties in selected closed-shell nuclei with masses ranging from $A=14$-208 for a set of chiral interactions. In this work we consider the \magicint{} interaction from Ref.~\cite{Hebe11fits}, the \deltago{} interaction with cutoff $\Lambda = 394 \, \MeV$ from Ref.~\cite{Jiang:2020the} as well as the recently developed \emarthuis{} interaction from Ref.~\cite{Arthuis2024}.
The two- and three-nucleon interactions are expanded in spherical harmonic oscillator states consisting of 15 major shells, including \NNN{} contributions at the normal-ordered three-body level. To ensure convergence in heavy systems the number of three-body configurations is allowed up to $E^{(3)}_\text{max}=24$ by exploiting the developments in Ref.~\cite{Miyagi2021}. The underlying HO frequency is varied between $\hbar \Omega = 10$--$16 \, \MeV$ and is taken from the minimum IMSRG ground-state energy for a given set of frequency values. The MBPT and IMSRG results for closed-shell nuclei used in this study are taken from Ref.~\cite{Arthuis2024}.

\subsection{Posterior distributions}

\begin{figure}[t!]
    \centering
    \includegraphics[width=\columnwidth]{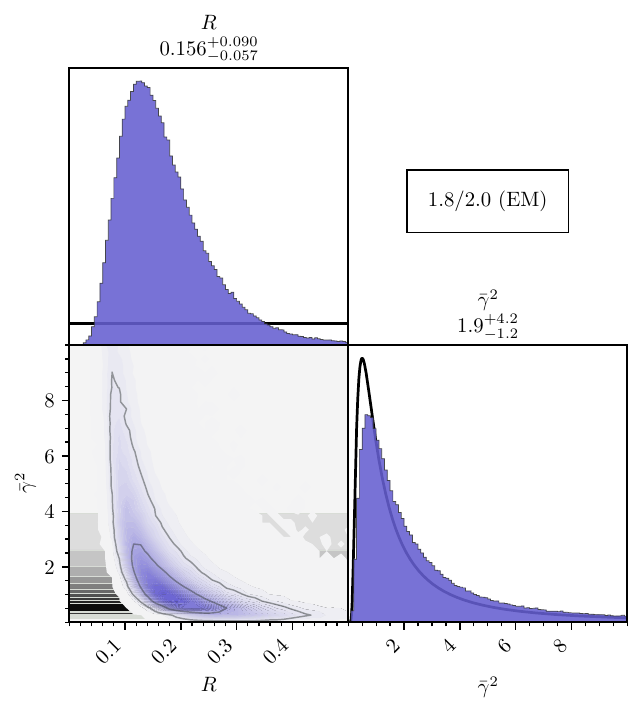}
    \caption{Priors (black lines/distributions) and posteriors (purple distributions) for $R$ and $\gbar^2$ obtained from the MBPT results with the \magicint{} interaction. The ranges for around the median of $R$ and $\gbar^2$ are given at the 68\% credibility level. The details of the inference are given in the text.}
    \label{fig:R_gamma2_posterior_magic}
\end{figure}

Using the theoretical values of $E_\text{HF}$, $\pte{2}$, and $\pte{3}$ for the given prior distributions, we construct a posterior distribution as described in Sec.~\ref{sec:stat}. For the inference we only use the nuclei $^{16}$O, $^{48}$Ca, and $^{132}$Sn in order to minimize the influence of correlations, since---as previously mentioned---the MBPT ratios are largely constant across different nuclei and we do not explicitly account for such correlations here. Including all available data would therefore lead to artificially narrow posterior distributions. We have verified that our results do not substantially change if we vary the number of nuclei in the inference between 2--5.
Figure~\ref{fig:R_gamma2_posterior_magic} shows the marginalized posterior distributions for $R$ and $\bar{\gamma}^2$ as well as their joint distribution for the results using the \magicint{} interaction. The prior distributions are additionally shown in the background.
The posterior distribution of $R$ reveals a pronounced peak at $R \approx 0.15$ with very little weight for values $R > 0.4$. The MBPT results from the three training nuclei clearly support a rapid suppression of higher-order corrections or, equivalently, small values for $R$. This is again directly related to the softness of the \magicint{} interaction.

In principle, predictions will depend on the choice of the prior distributions. To verify that our results are robust we have repeated our analysis using different prior choices, and posterior distributions for different priors for $\gbar^2$ are provided in App.~\ref{app:priors}. The overall sensitivity to the choice of prior is, while noticeable, small.

\subsection{Uncertainties in closed-shell nuclei}
\label{sec:closed-shell}

Next, we turn to the construction of uncertainty bands for the ground-state energies of atomic nuclei.
For the training we use a representative set of three closed-shell nuclei, \ie{}, \elem{16}{O}, \elem{48}{Ca} and \elem{132}{Sn}. We then use the obtained values of the MBPT error model, \ie{}, $\gbar^2$ and $R$, and draw a large number of truncation errors from the corresponding probability distribution for each nucleus.
From this we can construct credibility intervals for the predictions of all nuclei.
We give results for the energy per particle to compare the results over this broad range of mass number. A similar uncertainty for $E/A$ for all nuclei also reflects the size extensivity of MBPT.

\begin{figure}[t!]
    \centering
    \includegraphics[width=\linewidth]{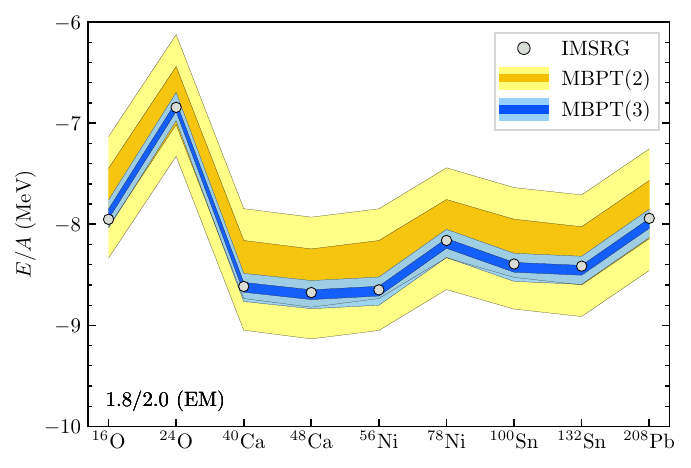}
    \caption{PPDs for the ground-state energy per nucleon of selected closed-shell nuclei at second and third order, MBPT(2) and MBPT(3) respectively, compared to IMSRG(2) results. Results are shown for the \magicint{} interaction. The dark (light) shaded areas indicate 68\% (90\%) credibility intervals.}
    \label{fig:ppd_magic}
\end{figure}

Figure~\ref{fig:ppd_magic} shows the PPDs at $68\%$ (dark shading) and $90\%$ (light shading) credibility levels for the second-order MBPT(2) (yellow) and third-order MBPT(3) (blue) ground-state energies for the \magicint{} interaction.
The second-order bands are sizable in all cases with uncertainties per particle of up to $1.0 \, \MeV$ at the $90\%$ level. This uncertainty is significantly reduced, to below $0.2\,\MeV$, when incorporating third-order corrections in heavier systems.
For comparison we include nonperturbative IMSRG(2) results in the same model space and with the same interaction. Due to the IMSRG resummation of diagrams the corresponding ground-state energies are expected to be much more accurate. We observe that in all cases the MBPT(3) results agree with the IMSRG calculations at the $90\%$ level.

We have repeated our analysis for different sets of nuclei. In particular we reduced the set of training systems to two nuclei without significantly reducing the quality of the results, albeit with somewhat wider uncertainty bands. This insensitivity is due to the strong correlations of $R$ values for different mass number based on the size-extensivity of the HF and MBPT energies. Reducing the training data further to a single nucleus, however, results in significantly wider bands. In the opposite extreme, \ie{}, including all 37 nuclei from Fig.~\ref{fig:mbptratios} in the inference, results in very narrow uncertainty bands due to the large number of strongly correlated data, which is treated as uncorrelated.

\subsection{Empirical coverage}

\begin{figure*}[t!]
    \centering
    \includegraphics[width=0.9\linewidth]{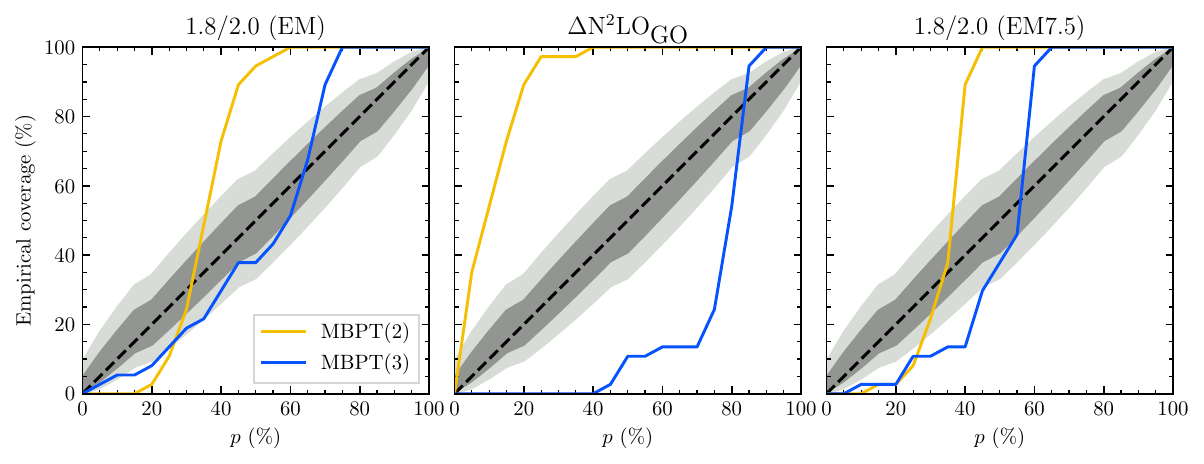}
    \caption{Empirical coverage (``weather'') plot corresponding to the predictions described in Fig.~\ref{fig:ppd_magic} (left panel), as well as analogous results obtained for the same nuclei calculated from the \deltago{} (center panel) and \emarthuis{} interactions (right panel). Values above the diagonal indicate too conservative UQ estimates and vice versa, with an ideal result following the diagonal. Results falling within the dark (light) gray area are consistent with the ideal result at the 68\% (95\%) confidence level given the amount of data we are comparing to (here the IMSRG results). }
    \label{fig:weatherman}
\end{figure*}

Based on our uncertainty estimates for the MBPT results we perform an analysis of the empirical coverage. The results are shown in Fig.~\ref{fig:weatherman}. An empirical coverage plot compares the expected coverage of the uncertainty bands with respect to some validation data with the observed (\ie{}, empirical) coverage obtained in practice. For example, a $p=50\%$ credibility interval (indicated on the $x$ axis) is expected to overlap with the validation data 50\% of the time; the observed overlap is shown on the $y$ axis.

Ideally the empirical coverage should fall on the diagonal, as indicated by the dashed black line in Fig.~\ref{fig:weatherman}. 
Values above this line correspond to a too conservative error estimate while values below indicate an underestimation of the associated error.
In general a too conservative error is preferable to an underestimated one in line with the precautionary principle.
Figure~\ref{fig:weatherman} shows the empirical coverage for our MBPT(2) and MBPT(3) results for the 37 closed-shell nuclei from Fig.~\ref{fig:mbptratios}. For the \magicint{} interaction shown in the left panel, the results indicate that the error model works largely as expected for the predictions at third order, but is conservative at second order. The figure also includes gray bands describing confidence intervals that indicate whether the obtained result is compatible with the ideal result~\cite{Furn15uncert}. If an empirical coverage falls within the dark (light) gray band, this should be interpreted as the empirical coverage being consistent with the ideal result at the 68\% (95\%) confidence level. In effect, these bands represent a ``blurred'' ideal diagonal that reflect the fact that we have a finite amount of data to compare to. Note that the gray bands assume that the comparison data is uncorrelated, which is not the case here; the gray bands are therefore too narrow~\cite{Melendez:2019izc}, as accounting for correlations would decrease the effective number of data. However, accounting for this correlation is nontrivial just as in the error model itself.

\subsection{Interaction sensitivity}
\label{sec:int_sens}

We continue our discussion by performing the uncertainty analysis for other chiral interactions and compare the size of the MBPT uncertainty to the results with the \magicint{} interaction.
To this end, we consider the \deltago{} interaction with cutoff $\Lambda = 394 \, \MeV$ from Ref.~\cite{Jiang:2020the}, and the \emarthuis{} interaction from Ref.~\cite{Arthuis2024} which features a refitted value of the three-nucleon coupling $c_D=7.5$, which leads to an improved description of charge radii compared to \magicint{}.

\begin{figure*}[t!]
    \centering
    \includegraphics[width=0.45\linewidth]{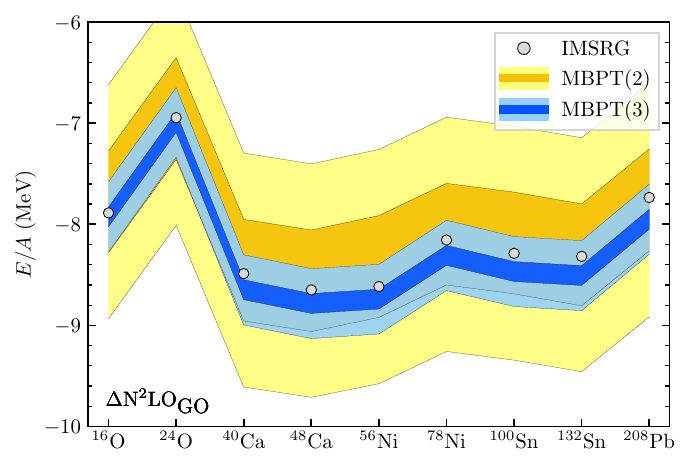}
    \includegraphics[width=0.45\linewidth]{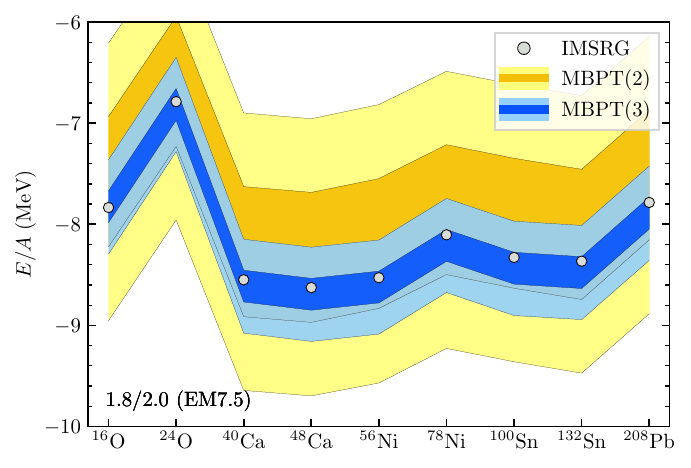}
    \caption{PPDs for the ground-state energy per nucleon of selected closed-shell nuclei as in Fig.~\ref{fig:ppd_magic} but for the \deltago{} (left panel) and \emarthuis{} interactions (right panel).}
    \label{fig:intcomp}
\end{figure*}

For comparing different interactions, we study the uncertainty bands for the same closed-shell nuclei as in Sec.~\ref{sec:closed-shell} using the same model-space parameters as for \magicint{}.
In Fig.~\ref{fig:intcomp} the uncertainties are shown for the \deltago{} (left panel) and \emarthuis{} interactions (right panel). Overall we observe increased uncertainties compared to the \magicint{} interaction, both at second and third order.
The width of the bands approximately doubles at both the $68\%$ and $90\%$ credibility levels.
Our results indicate that the \magicint{} interaction is thus more perturbative as indicated for instance by a larger HF energy and smaller MBPT corrections.
The \emarthuis{} interaction results in larger uncertainties compared to the \deltago{} interaction, with a $0.2 \, \MeV$ uncertainty of MBPT(3) already at the $68\%$ credibility level. This suggests that the \deltago{} interaction is more perturbative compared to \emarthuis{}.
Still, our uncertainty bands are in good agreement with the corresponding IMSRG(2) values that serve as reference in the absence of an exact many-body solution.

\subsection{Nuclear matter results}
\label{sec:nuclearmatter}

\begin{figure}[t!]
    \centering
    \includegraphics[width=\columnwidth]{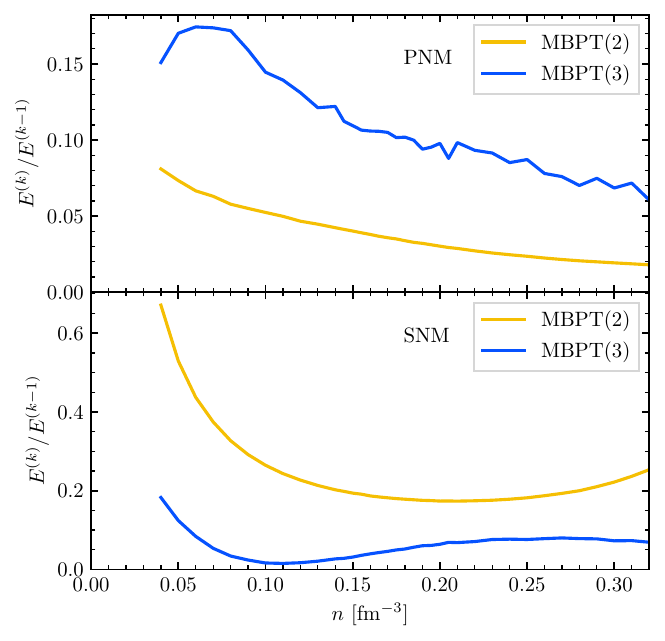}
    \caption{Ratios $\pte{2}/\Eref$ and $\pte{3}/\pte{2}$ as a function of density for pure neutron matter (PNM, top panel) and symmetric nuclear matter (SNM, bottom panel) from Ref.~\cite{Alp:2025wjn} obtained for the \magicint{} interaction.}
    \label{fig:NM_ratios}
\end{figure}

It is also possible to include nuclear matter results in the inference of many-body uncertainties. In Ref.~\cite{Alp:2025wjn} MBPT contributions up to third order have been computed based on the same interactions as those discussed for nuclei in Secs.~\ref{sec:closed-shell} and \ref{sec:int_sens}. Figure~\ref{fig:NM_ratios} shows the ratios of energy contributions at different MBPT orders for neutron matter (PNM) and symmetric nuclear matter (SNM) up to $2 n_0$, with saturation density $n_0 = 0.16$\,fm$^{-3}$. We find that including the SNM results at two representative densities, $n=0.5 n_0$ and $n=n_0$ (see Fig.~\ref{fig:mbptratios}), in the inference has no sizable impact on the inferred MBPT uncertainties. This is not too surprising because the ratios $\pte{2}/\Eref$ and $\pte{3}/\pte{2}$ are very similar in size to the ones we find in nuclei, as shown in Fig.~\ref{fig:mbptratios}. On the other hand, the MBPT convergence for PNM is qualitatively different and including PNM ratios significantly changes the MBPT uncertainty bands for nuclei.
This indicates that the MBPT expansion for PNM behaves different to the expansion for finite nuclei or SNM, and, hence, that the proposed error model has challenges capturing the MBPT convergence for nuclei and PNM simultaneously.

\subsection{Limitations of the error model}
\label{sec:limitations}

We now turn to a chiral interaction that is expected to be significantly less perturbative than the interactions discussed in the previous sections. The \nnlosat{} interaction from Ref.~\cite{Ekst15sat} is known to accurately describe ground-state properties of medium-mass nuclei, but is less often applied to heavier systems due to its notoriously slow convergence with respect to model-space size.
In Table~\ref{tab:mbptcomp} we provide a comparison of the many-body results for \elem{48}{Ca} based on the \magicint{} and \nnlosat{} interactions at different orders in the MBPT expansion.
While the two interactions yield comparable values for the total IMSRG(2) energy, the individual many-body contributions at different orders are very different in size. In case of the \magicint{} interaction $73\%$ of the total energy originates from the HF contribution, compared to only 30 \% for the \nnlosat{} interaction. At second order the picture is reversed. These numbers imply that for \nnlosat{} the HF state is a poorer reference state and we can expect the MBPT expansion to deteriorate. Interestingly, the third-order energy correction is comparable in size for both interactions; the third-order ratio $\pte{3}/\pte{2}$ is even a factor two smaller for the harder \nnlosat{} interaction. 
However, the smallness of $\pte{3}$ is accidental in this case, as one can see from the cancellation of the individually large particle-particle (pp) and particle-hole (ph) contributions at third order. This shows that one has to be careful with the size of MBPT uncertainties for hard interactions such as \nnlosat{}. Applying our statistical model to \nnlosat{} yields PPDs for medium-mass nuclei that are more than twice as wide compared to \magicint{}. However, we stress that knowledge of higher MBPT orders is likely necessary to accurately describe its convergence pattern.

To see the effect of an apparently divergent expansion, we have also studied a fictitious interaction based on the results of \nnlosat{} where we scaled up the third-order correction such that $\pte{3}/\pte{2} = \pte{2}/\Eref \approx 2.3$. In this case our error model finds $R > 1$, as expected, and the resulting PPDs are infinitely wide.

\begin{table}[t!]
\centering
\renewcommand{\arraystretch}{1.6}
\begin{tabular*}{\linewidth}{@{\extracolsep{\fill}}l S[table-format=2.4] S[table-format=3.3]}  
    \hline \hline
    & \multicolumn{1}{c}{\magicint{}} & \multicolumn{1}{c}{\nnlosat{}} \\ \hline 
    $E_\text{HF}$ [\MeV]& -303.581 & -122.180 \\
    \pte{2} [\MeV] & -105.926 & -281.942 \\
    \pte{3} [\MeV] & -7.904 & -11.369 \\
    \hline
    $E^{(3)}_\text{pp}$ [\MeV] & 0.932 & 12.733 \\
    $E^{(3)}_\text{hh}$ [\MeV] & -5.103 & -8.941 \\
    $E^{(3)}_\text{ph}$ [\MeV] & -3.733 & -15.162 \\
    \hline
    $\pte{2}/\Eref$ & 0.350 & 2.308 \\
    $\pte{3}/\pte{2}$ & 0.075 & 0.040 \\
    \hline
    IMSRG(2) [\MeV] & -416.399 & -405.275 \\
    \hline \hline
\end{tabular*}
\caption{Comparison of many-body results for the \magicint{} and \nnlosat{} interactions for the ground-state energy of \elem{48}{Ca}. Results are performed for $e_\text{max}=14$, $E^{(3)}_\text{max}=24$, and $\hbar \Omega = 16\, \MeV$ at the normal-ordered two-body level~\cite{Arthuis2024,Heinz2024calcium}.}
\label{tab:mbptcomp}
\end{table}
 
\section{Conclusion and outlook}
\label{sec:outlook}

We have developed a framework that allows for a statistical modeling of many-body uncertainties in \ai{} nuclear theory.
We considered MBPT as the simplest case of a correlation expansion, which allows for a straightforward modeling of the uncertainties from the many-body expansion. Within our Bayesian framework the model parameters were inferred from MBPT calculations of closed-shell nuclei ranging from $A=14$--$208$.
Our investigations showed only a mild sensitivity to the employed prior distribution of the model parameters, although only low-order MBPT results are included, which can lead to some prior influence especially if the priors are poorly chosen. As expected, the resulting uncertainties show a strong sensitivity to the employed interaction, in particular to its softness. We do not explicitly account for the fact that our input data is strongly correlated, and instead opt to only include a small subset of the available data in the inference. An important next step will be a proper treatment of correlations between the results for different nuclei.

The availability of robust models for MBPT uncertainties enables the estimate of combined many-body and interaction uncertainties in the future. In addition, the presented framework will allow the study of MBPT uncertainties for open-shell nuclei based on more general reference states, see Refs.~\cite{Tich17NCSM-MCPT,Tichai18BMBPT,Frosini2021mrIII,Demol20BMBPT,Demol2025pna}. This would also allow insights to correlated uncertainties along isotopic chains.

In addition, we included MBPT results at typical densities of symmetric nuclear matter and found that these results only have a mild effect of the Bayesian inference. However, the analysis of pure neutron matter showed a different convergence pattern than symmetric matter. In addition, nucleonic matter becomes nonperturbative at very low densities. Exploring MBPT uncertainties for the different conditions in nucleonic matter within our Bayesian framework will be an important next step, and will also require understanding correlations between different densities (and temperatures).

An in-depth understanding of MBPT truncations provides a first step to extend the modeling to more complicated many-body expansions that involve an infinite resummation of classes of diagrams such as the in-medium similarity renormalization group, coupled-cluster theory, or Green's function approaches, that provide the workhorses for \ai{} calculations of medium-mass and heavy nuclei.

\section*{Code and data availability}
The code and data that support the findings of this article are openly available \cite{zenodorepo}.

\section*{Acknowledgements}

We thank R.J.~Furnstahl and Z.~Li for useful discussions, P.~Arthuis, M.~Heinz and Z.~Li for sharing MBPT and IMSRG results, T. Miyagi for sharing MBPT(4) results, and Y.~Dietz and F.~Alp for sharing MBPT results for nuclear matter. 
This work was supported in part by the European Research Council (ERC) under the European Union's Horizon 2020 research and innovation programme (Grant Agreement No.~101020842) and under the European Union's Horizon Europe research and innovation programme (Grant Agreement No.~101162059).

\appendix

\section{Robustness to prior hyperparameters}
\label{app:priors}

\begin{figure*}[t!]
    \centering
    \includegraphics[width=0.49\linewidth]{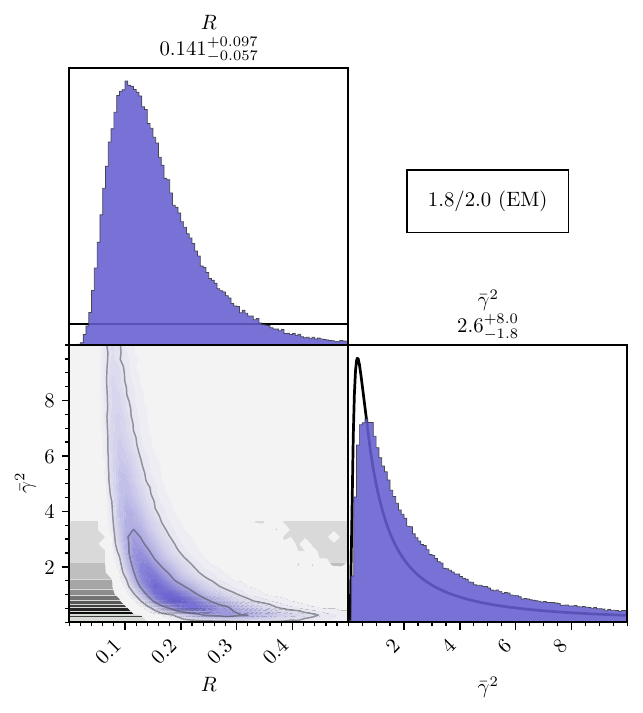}
    \includegraphics[width=0.49\linewidth]{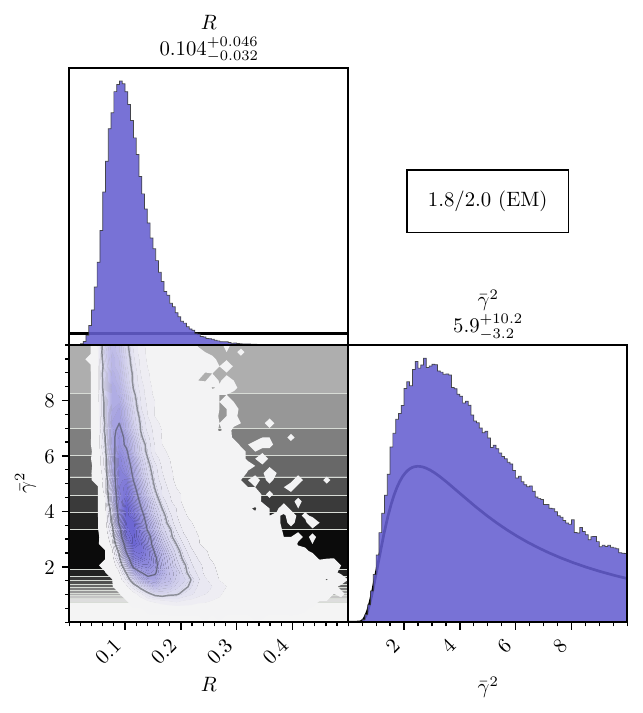}
    \caption{Posteriors with $\prob{\gbar^2 | I} = \ig \left(0.5, 0.5\right)$ (left panel) and $\prob{\gbar^2 | I} = \ig \left(1, 5\right)$ (right panel).}
    \label{fig:posterior_sensitivity}
\end{figure*}

\begin{figure*}[t!]
    \centering
    \includegraphics[width=0.49\linewidth]{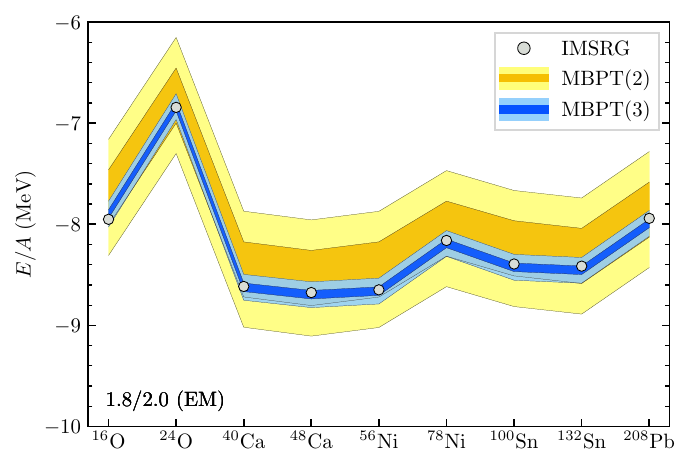}
    \includegraphics[width=0.49\linewidth]{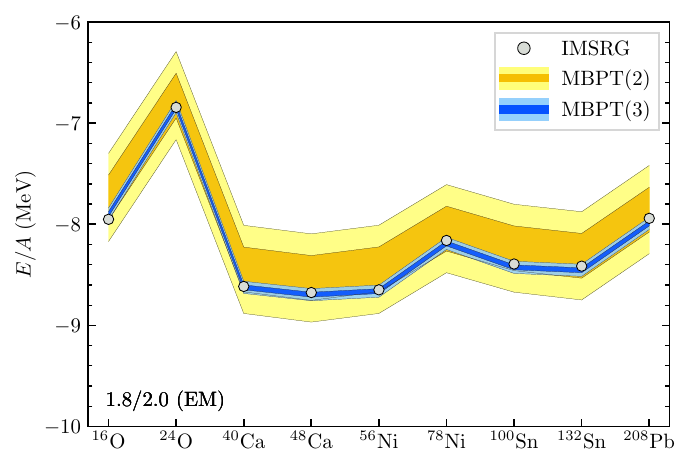}
    \caption{PPDs with $\prob{\gbar^2 | I} = \ig \left(0.5, 0.5\right)$ (left panel) and $\prob{\gbar^2 | I} = \ig \left(1, 5\right)$ (right panel).}
    \label{fig:ppd_sensitivity}
\end{figure*}

\begin{figure*}[t!]
    \centering
    \includegraphics[width=0.46\linewidth]{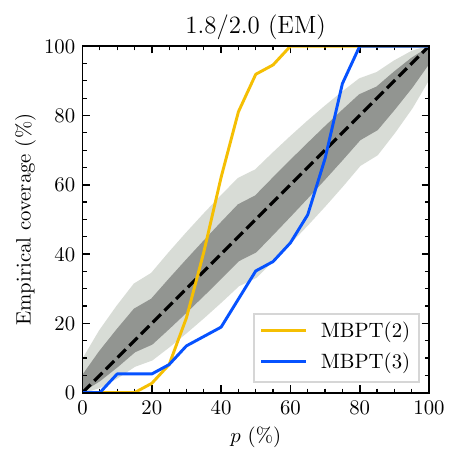}
    \vspace{5mm}
    \includegraphics[width=0.46\linewidth]{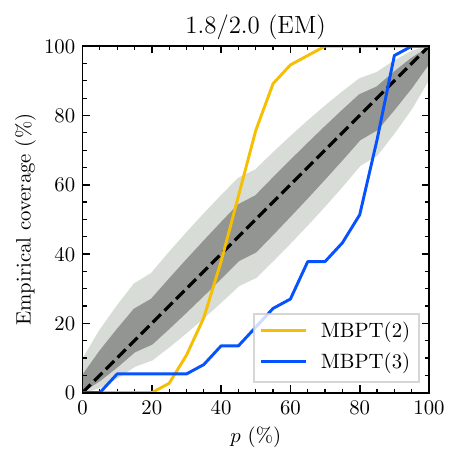}
    \caption{Empirical coverages with $\prob{\gbar^2 | I} = \ig \left(0.5, 0.5\right)$ (left panel) and $\prob{\gbar^2 | I} = \ig \left(1, 5\right)$ (right panel).}
    \label{fig:weatherman_sensitivity}
\end{figure*}

Here we show figures that allow us to judge the robustness of our results with regard to our prior choices. Figures~\ref{fig:posterior_sensitivity}, \ref{fig:ppd_sensitivity}, and \ref{fig:weatherman_sensitivity} show posterior, PPD, and empirical coverage plots, respectively, for reanalyses with modified priors for $\gbar^2$. The modified priors (shown in Fig.~\ref{fig:gamma2_prior}) are $\prob{\gbar^2 | I} = \ig \left(0.5, 0.5\right)$, which has more strength at small values of $\gbar^2$ while penalizing larger values, and $\prob{\gbar^2 | I} = \ig \left(1, 5\right)$ that substantially penalizes small values. The results obtained with the former prior are virtually identical to our main results. While the values for $R$ and $\gbar^2$ inferred with the latter prior differ substantially from our main results, this choice only leads to somewhat narrower credibility intervals and lower empirical coverages in the end. The increased value for $\gbar^2$ appears to be compensated by a smaller $R$ value. Given the rather drastic change of prior and our limited amount of data, we judge our results to be fairly robust with regards to the choice of prior.

\bibliography{bibliography}

@article{Maris:2020qne,
    author = "Maris, P. and Epelbaum, E. and Furnstahl, R.J. and Golak, J. and Hebeler, K. and H{\"u}ther, T. and Kamada, H. and Krebs, H. and Mei{\ss}ner, U.-G. and Melendez, J. A. and Nogga, A. and Reinert, P. and Roth, R. and Skibi\'nski, R. and Soloviov, V. and Topolnicki, K. and Vary, J. P. and Volkotrub, Yu. and Wita{\l}a, H. and Wolfgruber, T.",
    title = "{Light nuclei with semilocal momentum-space regularized chiral interactions up to third order}",
    doi = "10.1103/PhysRevC.103.054001",
    journal = "Phys. Rev. C",
    volume = "103",
    number = "5",
    pages = "054001",
    year = "2021"
}

@article{Tews:2024owl,
    author = {Tews, Ingo and Somasundaram, Rahul and Lonardoni, Diego and G{\"o}ttling, Hannah and Seutin, Rodric and Carlson, Joseph and Gandolfi, Stefano and Hebeler, Kai and Schwenk, Achim},
    title = "{Neutron matter from local chiral effective field theory interactions at large cutoffs}",
    doi = "10.1103/r314-6r62",
    journal = "Phys. Rev. Res.",
    volume = "7",
    number = "3",
    pages = "033024",
    year = "2025"
}

@article{Marino:2024tfp,
    author = "Marino, Francesco and Jiang, Weiguang and Novario, Samuel J.",
    title = "{Diagrammatic ab initio methods for infinite nuclear matter with modern chiral interactions}",
    doi = "10.1103/PhysRevC.110.054322",
    journal = "Phys. Rev. C",
    volume = "110",
    number = "5",
    pages = "054322",
    year = "2024"
}

@article{Sun:2024iht,
    author = {Sun, Z. H. and Ekstr{\"o}m, A. and Forss{\'e}n, C. and Hagen, G. and Jansen, G. R. and Papenbrock, T.},
    title = "{Multiscale Physics of Atomic Nuclei from First Principles}",
    doi = "10.1103/PhysRevX.15.011028",
    journal = "Phys. Rev. X",
    volume = "15",
    number = "1",
    pages = "011028",
    year = "2025"
}

@article{Miyagi:2023zvv,
    author = "Miyagi, T. and Cao, X. and Seutin, R. and Bacca, S. and Garcia Ruiz, R. F. and Hebeler, K. and Holt, J. D. and Schwenk, A.",
    title = "{Impact of Two-Body Currents on Magnetic Dipole Moments of Nuclei}",
    doi = "10.1103/PhysRevLett.132.232503",
    journal = "Phys. Rev. Lett.",
    volume = "132",
    number = "23",
    pages = "232503",
    year = "2024"
}

@article{Cirigliano:2022rmf,
    author = "Cirigliano, V. and Davoudi, Z. and Engel, J. and Furnstahl, R. J. and Hagen, G. and Heinz, U. and Hergert, H. and Horoi, M. and Johnson, C.W. and Lovato, A. and Mereghetti, E. and Nazarewicz, W. and Nicholson, A. and Papenbrock, T. and Pastore, S. and Plumlee, M. and Phillips, D. R. and Shanahan, P. E. and Stroberg, S. R. and Viens, F. and Walker-Loud, A. and Wendt, K. A. and Wild, S. M.",
    title = "{Towards precise and accurate calculations of neutrinoless double-beta decay}",
    doi = "10.1088/1361-6471/aca03e",
    journal = "J. Phys. G",
    volume = "49",
    number = "12",
    pages = "120502",
    year = "2022"
}

@article{Belley:2023lec,
    author = "Belley, A. and Yao, J. M. and Bally, B. and Pitcher, J. and Engel, J. and Hergert, H. and Holt, J. D. and Miyagi, T. and Rodriguez, T. R. and Romero, A. M. and Stroberg, S. R. and Zhang, X.",
    title = "{Ab~initio Uncertainty Quantification of Neutrinoless Double-Beta Decay in $^{76}$Ge}",
    doi = "10.1103/PhysRevLett.132.182502",
    journal = "Phys. Rev. Lett.",
    volume = "132",
    number = "18",
    pages = "182502",
    year = "2024"
}

@article{Alp:2025wjn,
    author = "Alp, Faruk and Dietz, Yannick and Hebeler, Kai and Schwenk, Achim",
    title = "{Equation~of state and Fermi liquid properties of dense matter based on chiral effective field theory interactions}",
    doi = "10.1103/ls3l-dn1y",
    journal = "Phys. Rev. C",
    volume = "112",
    number = "5",
    pages = "055802",
    year = "2025"
}

@article{Arthuis2024,
    author = "Arthuis, P. and Hebeler, K. and Schwenk, A.",
    title = "{Neutron-rich nuclei and neutron skins from chiral low-resolution interactions}",
    eprint = "2401.06675",
    archivePrefix = "arXiv",
    year = "2024",
    journal = ""
}

@article{Soma20SCGF,
    author = "Som{\`a}, V. and Navr{\'a}til, P. and Raimondi, F. and Barbieri, C. and Duguet, T.",
    title = "{Novel chiral Hamiltonian and observables in light and medium-mass nuclei}",
    doi = "10.1103/PhysRevC.101.014318",
    journal = "Phys. Rev. C",
    volume = "101",
    pages = "014318",
    year = "2020"
}

@article{Herg20review,
    author={Hergert, Heiko},   
    title={A {G}uided {T}our of ab initio {N}uclear {M}any-{B}ody {T}heory},      
	journal={Front. Phys.},      
    volume={8},      
    pages={379},     
	year={2020},      
	url-={https://www.frontiersin.org/article/10.3389/fphy.2020.00379},       
	doi={10.3389/fphy.2020.00379},      
	ISSN-={2296-424X},   
}

@InBook{Barb17SCGFlectnote,
  pages     = {571},
  title     = {Self-Consistent {G}reen's Function Approaches},
  publisher = {Springer International Publishing},
  year      = {2017},
  author    = {Barbieri, Carlo and Carbone, Arianna},
  editor    = {Hjorth-Jensen, Morten and Lombardo, Maria Paola and van Kolck, Ubirajara},
  address   = {Cham},
  isbn      = {978-3-319-53336-0},
  booktitle = {An Advanced Course in Computational Nuclear Physics: Bridging the Scales from Quarks to Neutron Stars},
  doi      = {10.1007/978-3-319-53336-0_11},
  url-      = {https://doi.org/10.1007/978-3-319-53336-0_11},
}

@article{Barr13PPNP,
      author         = "Barrett, B. R. and Navr{\'a}til, P. and Vary, J. P.",
      title          = "{Ab initio no core shell model}",
      journal        = "Prog. Part. Nucl. Phys.",
      volume         = "69",
      year           = "2013",
      pages          = "131",
      doi           = "10.1016/j.ppnp.2012.10.003",
      SLACcitation   = "%%CITATION = PPNPD,69,131;%%"
}

@Article{Carl15RMP,
  author  = {Carlson, J. and Gandolfi, S. and Pederiva, F. and Pieper, S. C. and Schiavilla, R. and Schmidt, K. E. and Wiringa, R. B.},
  title   = {{Quantum Monte Carlo methods for nuclear physics}},
  journal = {Rev. Mod. Phys.},
  year    = {2015},
  volume  = {87},
  pages   = {1067},
  doi    = {10.1103/RevModPhys.87.1067},
}

@article{Demol20BMBPT,
title = {Bogoliubov many-body perturbation theory under constraint},
journal = {Ann. Phys.},
volume = {424},
pages = {168358},
year = {2021},
issn = {0003-4916},
doi = {https://doi.org/10.1016/j.aop.2020.168358},
url = {https://www.sciencedirect.com/science/article/pii/S000349162030292X},
author = {P. Demol and M. Frosini and A. Tichai and V. Somà and T. Duguet},
}

@article{Demol2025pna,
	author = {{Demol, P.} and {Duguet, T.} and {Tichai, A.}},
	title = "{Ab initio Bogoliubov many-body perturbation theory: closed-form constraint on the average particle number}",
	DOI= "10.1140/epja/s10050-024-01480-7",
	url= "https://doi.org/10.1140/epja/s10050-024-01480-7",
	journal = {Eur. Phys. J. A},
	year = 2025,
	volume = 61,
	number = 1,
	pages = "16",
}

@article{Door2025ytterbium,
    title={Search for new bosons with ytterbium isotope shifts}, 
    author={Menno Door and Chih-Han Yeh and Matthias Heinz and Fiona Kirk and Chunhai Lyu and Takayuki Miyagi and Julian C. Berengut and Jacek Bieroń and Klaus Blaum and Laura S. Dreissen and Sergey Eliseev and Pavel Filianin and Melina Filzinger and Elina Fuchs and Henning A. Fürst and Gediminas Gaigalas and Zoltán Harman and Jost Herkenhoff and Nils Huntemann and Christoph H. Keitel and Kathrin Kromer and Daniel Lange and Alexander Rischka and Christoph Schweiger and Achim Schwenk and Noritaka Shimizu and Tanja E. Mehlstäubler},
    doi = "10.1103/PhysRevLett.134.063002",
    journal = "Phys. Rev. Lett.",
    volume = "134",
    number = "6",
    pages = "063002",
    year = "2025"
}

@article{Dris17MCshort,
      author         = "Drischler, C. and Hebeler, K. and Schwenk, A.",
      title          = "{Chiral {I}nteractions up to
                        {N}ext-to-{N}ext-to-{N}ext-to-{L}eading {O}rder and {N}uclear
                        {S}aturation}",
      journal        = "Phys. Rev. Lett.",
      volume         = "122",
      year           = "2019",
      number         = "4",
      pages          = "042501",
      doi            = "10.1103/PhysRevLett.122.042501",
      eprint-         = "1710.08220",
      SLACcitation   = "%%CITATION = ARXIV:1710.08220;%%"
}

@article{Ekst15sat,
      author            = {Ekstr\"om, A. and Jansen, G. R. and Wendt, K. A. and Hagen, G. and Papenbrock, T. and Carlsson, B. D. and Forss\'en, C. and Hjorth-Jensen, M. and Navr\'atil, P. and Nazarewicz, W.},
      title          = "{Accurate nuclear radii and binding energies from a
                        chiral interaction}",
      journal        = "Phys. Rev. C",
      volume         = "91",
      pages          = "051301(R)",
      doi            = "10.1103/PhysRevC.91.051301",
      year           = "2015",
      eprint-        = "1502.04682",
      SLACcitation   = "%%CITATION = ARXIV:1502.04682;%%",
}

@Article{Epel09RMP,
  author       = {E. Epelbaum and H.-W. Hammer and U.-G. Mei{\ss}ner},
  title        = {{Modern theory of nuclear forces}},
  journal      = {Rev. Mod. Phys.},
  year         = {2009},
  volume       = {81},
  pages        = {1773},
  doi         = {10.1103/RevModPhys.81.1773},
  eprint-      = {0811.1338},
  reportnumber = {HISKP-TH-08-18, FZJ-IKP-TH-2008-20},
  slaccitation = {%%CITATION = ARXIV:0811.1338;%%},
}

@article{Frosini2021mrII,
    author = "Frosini, Mikael and Duguet, Thomas and Ebran, Jean-Paul and Bally, Benjamin and Mongelli, Tobias and Rodr\'\i{}guez, Tom\'as R. and Roth, Robert and Som\`a, Vittorio",
    title = "{Multi-reference many-body perturbation theory for nuclei: II. Ab initio study of neon isotopes via PGCM and IM-NCSM calculations}",
    doi = "10.1140/epja/s10050-022-00693-y",
    journal = "Eur. Phys. J. A",
    volume = "58",
    number = "4",
    pages = "63",
    year = "2022"
}

@article{Frosini2021mrIII,
    author = "Frosini, Mikael and Duguet, Thomas and Ebran, Jean-Paul and Bally, Benjamin and Hergert, Heiko and Rodr\'\i{}guez, Tom\'as R. and Roth, Robert and Yao, Jiangming and Som\`a, Vittorio",
    title = "{Multi-reference many-body perturbation theory for nuclei: III. Ab initio calculations at second order in PGCM-PT}",
    doi = "10.1140/epja/s10050-022-00694-x",
    journal = "Eur. Phys. J. A",
    volume = "58",
    number = "4",
    pages = "64",
    year = "2022"
}

@article{Furn15uncert,
      author         = "Furnstahl, R. J. and Klco, N. and Phillips, D. R. and
                        Wesolowski, S.",
      title          = "{Quantifying truncation errors in effective field
                        theory}",
      journal        = "Phys. Rev. C",
      volume         = "92",
      year           = "2015",
      number         = "2",
      pages          = "024005",
      doi           = "10.1103/PhysRevC.92.024005",
      eprint-        = "1506.01343",
      SLACcitation   = "%%CITATION = ARXIV:1506.01343;%%"
}

@Article{Gold57mbpt,
  author    = {Goldstone, J.},
  title     = "{Derivation of the Brueckner many-body theory}",
  journal   = {Proc. Roy. Soc. A},
  year      = {1957},
  volume    = {239},
  number    = {1217},
  pages     = {267},
  issn      = {0080-4630},
  doi      = {10.1098/rspa.1957.0037},
  eprint-   = {http://rspa.royalsocietypublishing.org/content/239/1217/267.full.pdf},
  publisher = {The Royal Society},
  url-      = {http://rspa.royalsocietypublishing.org/content/239/1217/267},
}

@Article{Hage14RPP,
  author       = {Hagen, G. and Papenbrock, T. and Hjorth-Jensen, M. and Dean, D. J.},
  title        = {{Coupled-cluster computations of atomic nuclei}},
  journal      = {Rep. Prog. Phys.},
  year         = {2014},
  volume       = {77},
  number       = {9},
  pages        = {096302},
  doi         = {10.1088/0034-4885/77/9/096302},
  eprint-      = {1312.7872},
  slaccitation = {%%CITATION = ARXIV:1312.7872;%%},
}

@article{Hagen2022PCC,
  title = {Angular-momentum projection in coupled-cluster theory: Structure of $^{34}\mathrm{Mg}$},
  author = {Hagen, G. and Novario, S. J. and Sun, Z. H. and Papenbrock, T. and Jansen, G. R. and Lietz, J. G. and Duguet, T. and Tichai, A.},
  journal = {Phys. Rev. C},
  volume = {105},
  issue = {6},
  pages = {064311},
  numpages = {23},
  year = {2022},
  month = {Jun},
  publisher = {American Physical Society},
  doi = {10.1103/PhysRevC.105.064311},
  url = {https://link.aps.org/doi/10.1103/PhysRevC.105.064311}
}

@article{Hebe11fits,
	Author = {Hebeler, K. and Bogner, S. K. and Furnstahl, R. J. and Nogga, A. and Schwenk, A.},
	doi = {10.1103/PhysRevC.83.031301},
	Eprint- = {1012.3381},
	Journal = {Phys. Rev. C},
	Pages = {031301(R)},
	Slaccitation = {%%CITATION = ARXIV:1012.3381;%%},
	Title = {{Improved nuclear matter calculations from chiral low-momentum interactions}},
	Volume = {83},
	Year = {2011}}

@article{Hebe203NF,
    author = "Hebeler, K.",
    title = "{Three-Nucleon Forces: Implementation and Applications to Atomic Nuclei and Dense Matter}",
    doi = "10.1016/j.physrep.2020.08.009",
    journal = "Phys. Rep.",
    volume = "890",
    pages = "1--116",
    year = "2021"
}

@article{Heinz2024calcium,
  title = {Improved structure of calcium isotopes from ab initio calculations},
  author = {Heinz, M. and Miyagi, T. and Stroberg, S. R. and Tichai, A. and Hebeler, K. and Schwenk, A.},
  journal = {Phys. Rev. C},
  volume = {111},
  issue = {3},
  pages = {034311},
  numpages = {14},
  year = {2025},
  month = {Mar},
  publisher = {American Physical Society},
  doi = {10.1103/PhysRevC.111.034311},
  url = {https://link.aps.org/doi/10.1103/PhysRevC.111.034311}
}

@Article{Herg16PR,
  author       = {Hergert, H. and Bogner, S. K. and Morris, T. D. and Schwenk, A. and Tsukiyama, K.},
  title        = {{The In-Medium Similarity Renormalization Group: A Novel Ab Initio Method for Nuclei}},
  journal      = {Phys. Rep.},
  year         = {2016},
  volume       = {621},
  pages        = {165},
  doi         = {10.1016/j.physrep.2015.12.007},
  eprint-      = {1512.06956},
  slaccitation = {%%CITATION = ARXIV:1512.06956;%%},
}

@article{Stroberg2021,
  title = {Ab Initio Limits of Atomic Nuclei},
  author = {Stroberg, S. R. and Holt, J. D. and Schwenk, A. and Simonis, J.},
  journal = {Phys. Rev. Lett.},
  volume = {126},
  issue = {2},
  pages = {022501},
  numpages = {6},
  year = {2021},
  month = {Jan},
  publisher = {American Physical Society},
  doi = {10.1103/PhysRevLett.126.022501},
  url = {https://link.aps.org/doi/10.1103/PhysRevLett.126.022501}
}

@article{Hopp19medmass,
      author         = "Hoppe, J. and Drischler, C. and Hebeler, K. and Schwenk,
                        A. and Simonis, J.",
      title          = "{Probing chiral interactions up to
                        next-to-next-to-next-to-leading order in medium-mass
                        nuclei}",
      journal        = "Phys. Rev. C",
      volume         = "100",
      year           = "2019",
      number         = "2",
      pages          = "024318",
      doi            = "10.1103/PhysRevC.100.024318",
      eprint-         = "1904.12611",
      SLACcitation   = "%%CITATION = ARXIV:1904.12611;%%"
}

@article{Hu2021lead,
    author = {Hu, Baishan and Jiang, Weiguang and Miyagi, Takayuki and Sun, Zhonghao and Ekström, Andreas and Forssén, Christian and Hagen, Gaute and Holt, Jason D. and Papenbrock, Thomas and Stroberg, S. Ragnar and Vernon, Ian},
    title = "{Ab initio predictions link the neutron skin of ${}^{208}$Pb to nuclear forces}",
    journal = {Nat. Phys.},
    volume={18},
    page={1196–1200},
    url={https://doi.org/10.1038/s41567-022-01715-8},
    month = "12",
    year = "2021"
}

@article{Huth19chiralfam,
author = "Hüther, T. and Vobig, K. and Hebeler, K. and Machleidt, R. and Roth, R.",
title = "Family of chiral two- plus three-nucleon interactions for accurate nuclear structure studies",
journal = "Phys. Lett. B",
volume = "808",
pages = "135651",
year = "2020",
issn = "0370-2693",
doi = "https://doi.org/10.1016/j.physletb.2020.135651",
}

@article{Keller2023,
  title = {Nuclear Equation of State for Arbitrary Proton Fraction and Temperature Based on Chiral Effective Field Theory and a {Gaussian} Process Emulator},
  author = {Keller, J. and Hebeler, K. and Schwenk, A.},
  journal = {Phys. Rev. Lett.},
  volume = {130},
  issue = {7},
  pages = {072701},
  numpages = {6},
  year = {2023},
  month = {Feb},
  publisher = {American Physical Society},
  doi = {10.1103/PhysRevLett.130.072701},
  url = {https://link.aps.org/doi/10.1103/PhysRevLett.130.072701}
}

@article{Mach11PR,
	Author = {Machleidt, R. and Entem, D. R.},
	doi = {10.1016/j.physrep.2011.02.001},
	Eprint- = {1105.2919},
	Journal = {Phys. Rep.},
	Pages = {1},
	Slaccitation = {%%CITATION = ARXIV:1105.2919;%%},
	Title = {{Chiral effective field theory and nuclear forces}},
	Volume = {503},
	Year = {2011}}

@Article{Mele17bayerror,
  author         = {Melendez, J. A. and Wesolowski, S. and Furnstahl, R. J.},
  title          = {{Bayesian truncation errors in chiral effective field theory: nucleon-nucleon observables}},
  journal        = {Phys. Rev. C},
  year           = {2017},
  volume         = {96},
  number         = {2},
  pages          = {024003},
  archiveprefix- = {arXiv},
  doi           = {10.1103/PhysRevC.96.024003},
  eprint-        = {1704.03308},
  primaryclass-  = {nucl-th},
  slaccitation-  = {%%CITATION = ARXIV:1704.03308;%%},
}

@article{Miyagi2021,
  title = {Converged ab initio calculations of heavy nuclei},
  author = {Miyagi, T. and Stroberg, S. R. and Navr\'atil, P. and Hebeler, K. and Holt, J. D.},
  journal = {Phys. Rev. C},
  volume = {105},
  issue = {1},
  pages = {014302},
  numpages = {14},
  year = {2022},
  month = {Jan},
  publisher = {American Physical Society},
  doi = {10.1103/PhysRevC.105.014302},
  url = {https://link.aps.org/doi/10.1103/PhysRevC.105.014302}
}

@Misc{Miyagi25pc,
  author = {Miyagi, T.},
  note   = {{private communication (2025)}}
}

@article{Roth09ImTr,
	Author = {Roth, R.},
	doi = {10.1103/PhysRevC.79.064324},
	Eprint- = {0903.4605},
	Journal = {Phys. Rev. C},
	Pages = {064324},
	Slaccitation = {%%CITATION = ARXIV:0903.4605;%%},
	Title = {{Importance Truncation for Large-Scale Configuration Interaction Approaches}},
	Volume = {79},
	Year = {2009}}

@Book{Shav09MBmethod,
      author={Shavitt, Isaiah and Bartlett, Rodney J.},
      title={Many-Body Methods in Chemistry and Physics: MBPT and Coupled-Cluster Theory},
      series={Cambridge Molecular Science},
      collection={Cambridge Molecular Science},
      doi-={10.1017/CBO9780511596834},
      publisher={Cambridge University Press},
      place={Cambridge},
      year={2009}
}

@article{Stroberg2019,
author = {Stroberg, S. R. and Hergert, H. and Bogner, S. K. and Holt, J. D.},
title = {Nonempirical {I}nteractions for the {N}uclear {S}hell {M}odel: An {U}pdate},
journal = {Annu. Rev. Nucl. Part. Sci.},
volume = {69},
number = {1},
pages = {307},
year = {2019},
doi = {10.1146/annurev-nucl-101917-021120},
}

@article{Tich16HFMBPT,
      author         = "{A. Tichai} and Langhammer, J. and Binder,
                        S. and Roth, R.",
      title          = "{Hartree-Fock many-body perturbation theory for nuclear
                        ground-states}",
      journal        = "Phys. Lett. B",
      volume         = "756",
      year           = "2016",
      pages          = "283",
      doi            = "10.1016/j.physletb.2016.03.029",
      eprint-         = "1601.03703",
      options = {maxnames=10},
      SLACcitation   = "%%CITATION = ARXIV:1601.03703;%%"
}

@article{Tich17NCSM-MCPT,
    title = {Open-shell nuclei from No-Core Shell Model with perturbative improvement},
    journal = {Phys. Lett. B},
    volume = {786},
    pages = {448-452},
    year = {2018},
    issn = {0370-2693},
    doi = {https://doi.org/10.1016/j.physletb.2018.10.029},
    url = {https://www.sciencedirect.com/science/article/pii/S0370269318307986},
    author = {{A. Tichai} and Eskendr Gebrerufael and Klaus Vobig and Robert Roth},
}

@article{Tichai18BMBPT,
    author = "{A. Tichai} and Arthuis, P. and Duguet, T. and Hergert, H. and Somá, V. and Roth, R.",
    title = "{Bogoliubov many-body perturbation theory for open-shell nuclei}",
    eprint- = "1806.10931",
    doi = "10.1016/j.physletb.2018.09.044",
    journal = "Phys. Lett. B",
    volume = "786",
    pages = "195",
    year = "2018",
    options = {maxnames=10},
}

@article{Tichai2020review,
  author={{A. Tichai} and Roth, R. and Duguet, T.},   
  title={Many-{B}ody {P}erturbation {T}heories for {F}inite {N}uclei},      
  journal={Front. Phys.},      
  volume={8},      
  pages={164},     
  year={2020},      
  url-={https://www.frontiersin.org/article/10.3389/fphy.2020.00164},
  doi={10.3389/fphy.2020.00164},
}

@article{Tichai2024bcc,
title = {Towards heavy-mass ab initio nuclear structure: Open-shell {Ca}, {Ni} and {Sn} isotopes from {B}ogoliubov coupled-cluster theory},
journal = {Phys. Lett. B},
volume = {851},
pages = {138571},
year = {2024},
issn = {0370-2693},
doi = {https://doi.org/10.1016/j.physletb.2024.138571},
url = {https://www.sciencedirect.com/science/article/pii/S0370269324001291},
author = {A. Tichai and P. Demol and T. Duguet},
}

@Article{Weso16Bayes,
  author         = {Wesolowski, S. and Klco, N. and Furnstahl, R. J. and Phillips, D. R. and Thapaliya, A.},
  title          = {{Bayesian parameter estimation for effective field theories}},
  journal        = {J. Phys. G},
  year           = {2016},
  volume         = {43},
  number         = {7},
  pages          = {074001},
  archiveprefix- = {arXiv},
  doi           = {10.1088/0954-3899/43/7/074001},
  eprint-        = {1511.03618},
  primaryclass-  = {nucl-th},
  slaccitation-  = {%%CITATION = ARXIV:1511.03618;%%},
}

@article{Jiang:2020the,
    author = {Jiang, W. G. and Ekstr\"om, A. and Forss\'en, C. and Hagen, G. and Jansen, G. R. and Papenbrock, T.},
    title = "{Accurate bulk properties of nuclei from $A=2$ to $\infty$ from potentials with $\Delta$ isobars}",
    doi = "10.1103/PhysRevC.102.054301",
    journal = "Phys. Rev. C",
    volume = "102",
    number = "5",
    pages = "054301",
    year = "2020"
}

@article{Melendez:2019izc,
    author = "Melendez, J. A. and Furnstahl, R. J. and Phillips, D. R. and Pratola, M. T. and Wesolowski, S.",
    title = "Quantifying Correlated Truncation Errors in Effective Field Theory",
    doi = "10.1103/PhysRevC.100.044001",
    journal = "Phys. Rev. C",
    volume = "100",
    number = "4",
    pages = "044001",
    year = "2019"
}

@article{Wesolowski:2018lzj,
    author = "Wesolowski, S. and Furnstahl, R. J. and Melendez, J. A. and Phillips, D. R.",
    title = "{Exploring Bayesian parameter estimation for chiral effective field theory using nucleon\textendash{}nucleon phase shifts}",
    doi = "10.1088/1361-6471/aaf5fc",
    journal = "J. Phys. G",
    volume = "46",
    number = "4",
    pages = "045102",
    year = "2019"
}

@article{Wesolowski:2021cni,
    author = {Wesolowski, S. and Svensson, I. and Ekstr\"om, A. and Forss\'en, C. and Furnstahl, R. J. and Melendez, J. A. and Phillips, D. R.},
    title = "{Rigorous constraints on three-nucleon forces in chiral effective field theory from fast and accurate calculations of few-body observables}",
    doi = "10.1103/PhysRevC.104.064001",
    journal = "Phys. Rev. C",
    volume = "104",
    number = "6",
    pages = "064001",
    year = "2021"
}

@article{Svensson:2022kkj,
    author = {Svensson, Isak and Ekstr\"om, Andreas and Forss\'en, Christian},
    title = "{Bayesian estimation of the low-energy constants up to fourth order in the nucleon-nucleon sector of chiral effective field theory}",
    doi = "10.1103/PhysRevC.107.014001",
    journal = "Phys. Rev. C",
    volume = "107",
    number = "1",
    pages = "014001",
    year = "2023"
}

@article{Foreman_Mackey_2013,
   title="{\texttt{Emcee}: The MCMC Hammer}",
   volume={125},
   ISSN={1538-3873},
   url={http://dx.doi.org/10.1086/670067},
   DOI={10.1086/670067},
   number={925},
   journal={Publ. Astron. Soc. Pac.},
   publisher={IOP Publishing},
   author={Foreman-Mackey, Daniel and Hogg, David W. and Lang, Dustin and Goodman, Jonathan},
   year={2013},
   month=mar, pages={306}
}

@article{Drischler:2020hwi,
    author = "Drischler, C. and Furnstahl, R. J. and Melendez, J. A. and Phillips, D. R.",
    title = "{How Well Do We Know the Neutron-Matter Equation of State at the Densities Inside Neutron Stars? A Bayesian Approach with Correlated Uncertainties}",
    doi = "10.1103/PhysRevLett.125.202702",
    journal = "Phys. Rev. Lett.",
    volume = "125",
    number = "20",
    pages = "202702",
    year = "2020"
}

@Article{         harris2020array,
 title         = {Array programming with {NumPy}},
 author        = {Charles R. Harris and K. Jarrod Millman and St{\'{e}}fan J.
                 van der Walt and Ralf Gommers and Pauli Virtanen and David
                 Cournapeau and Eric Wieser and Julian Taylor and Sebastian
                 Berg and Nathaniel J. Smith and Robert Kern and Matti Picus
                 and Stephan Hoyer and Marten H. van Kerkwijk and Matthew
                 Brett and Allan Haldane and Jaime Fern{\'{a}}ndez del
                 R{\'{i}}o and Mark Wiebe and Pearu Peterson and Pierre
                 G{\'{e}}rard-Marchant and Kevin Sheppard and Tyler Reddy and
                 Warren Weckesser and Hameer Abbasi and Christoph Gohlke and
                 Travis E. Oliphant},
 year          = {2020},
 month         = sep,
 journal       = {Nature},
 volume        = {585},
 number        = {7825},
 pages         = {357--362},
 doi           = {10.1038/s41586-020-2649-2},
 publisher     = {Springer Science and Business Media {LLC}},
 url           = {https://doi.org/10.1038/s41586-020-2649-2}
}

@article{Millican:2024yuz,
    author = "Millican, P. J. and Furnstahl, R. J. and Melendez, J. A. and Phillips, D. R. and Pratola, M. T.",
    title = "{Assessing correlated truncation errors in modern nucleon-nucleon potentials}",
    doi = "10.1103/PhysRevC.110.044002",
    journal = "Phys. Rev. C",
    volume = "110",
    number = "4",
    pages = "044002",
    year = "2024"
}

@article{Svensson:2023twt,
    author = {Svensson, Isak and Ekstr{\"o}m, Andreas and Forss{\'e}n, Christian},
    title = "{Inference of the low-energy constants in {\ensuremath{\Delta}}-full chiral effective field theory including a correlated truncation error}",
    doi = "10.1103/PhysRevC.109.064003",
    journal = "Phys. Rev. C",
    volume = "109",
    number = "6",
    pages = "064003",
    year = "2024"
}

@article{Svensson:2021lzs,
    author = {Svensson, Isak and Ekstr{\"o}m, Andreas and Forss{\'e}n, Christian},
    title = "{Bayesian parameter estimation in chiral effective field theory using the Hamiltonian Monte Carlo method}",
    doi = "10.1103/PhysRevC.105.014004",
    journal = "Phys. Rev. C",
    volume = "105",
    number = "1",
    pages = "014004",
    year = "2022"
}

@article{Jiang:2022tzf,
    author = {Jiang, W. G. and Forss{\'e}n, C. and Dj{\"a}rv, T. and Hagen, G.},
    title = "{Nuclear-matter saturation and symmetry energy within {\ensuremath{\Delta}}-full chiral effective field theory}",
    doi = "10.1103/PhysRevC.109.L061302",
    journal = "Phys. Rev. C",
    volume = "109",
    number = "6",
    pages = "L061302",
    year = "2024"
}

@article{Armstrong:2025tza,
    author = "Armstrong, Cassandra L. and Giuliani, Pablo and Godbey, Kyle and Somasundaram, Rahul and Tews, Ingo",
    title = "{Emulators for Scarce and Noisy Data: Application to Auxiliary-Field Diffusion Monte~Carlo for Neutron Matter}",
    reportNumber = "LA-UR-24-27320",
    doi = "10.1103/9928-wyjm",
    journal = "Phys. Rev. Lett.",
    volume = "135",
    number = "14",
    pages = "142501",
    year = "2025"
}

@software{zenodorepo,
  author       = {Svensson, Isak and
                  Tichai, Alexander and
                  Hebeler, Kai and
                  Schwenk, Achim},
  title        = {{Reproduction package: A Bayesian approach for
                   many-body uncertainties in nuclear structure:
                   Many-body perturbation theory for finite nuclei
                  }},
  month        = sep,
  year         = 2025,
  publisher    = {Zenodo},
  version      = {v1.0},
  doi          = {10.5281/zenodo.17192609},
  url          = {https://doi.org/10.5281/zenodo.17192609},
  swhid        = {swh:1:dir:5c1a9dce4bc4edd87833150d5ae17feffacf10bb
                   ;origin=https://doi.org/10.5281/zenodo.17192608;vi
                   sit=swh:1:snp:551dd70f775d6618062a811eba9dde51b477
                   09ae;anchor=swh:1:rel:c0dedccda5a0fc47fbf5b53b3541
                   534ba5bb2ffe;path=svisak-manybody\_uncertainties-
                   cc24e24
                  },
}

\end{document}